\documentclass[12pt]{iopart}

\usepackage{bm}
\usepackage{graphicx}
\usepackage{iopams}

\begin{document}
\paper{Coherent electron dynamics in a two-dimensional random system with mobility edges}

\author{F A B F de Moura$^1$, M L Lyra$^1$, F Dom\'{\i}nguez-Adame$^2$ and\\
V A Malyshev$^{3,4}$\footnote{$^4$ On leave from V.\ A.\ Fock
Institute of Physics, Saint-Petersburg State University, 198904
Saint-Petersburg, Russia.}}

\address{$^1$ Instituto de F\'{\i}sica, Universidade Federal de
Alagoas, Macei\'{o} AL 57072-970, Brazil}
\address{$^2$ GISC, Departamento de F\'{\i}sica de Materiales, Universidad
Complutense, E-28040 Madrid, Spain}

\address{$^3$ Institute for Theoretical Physics and Materials
Science Centre, University of Groningen, Nijenborgh 4, 9747 AG
Groningen, The Netherlands}

\begin{abstract}

We study numerically the dynamics of a one-electron wave packet in
a two-dimensional random lattice with long-range correlated
diagonal disorder in the presence of a uniform electric field. The
time-dependent Schr\"{o}dinger equation is used for this purpose.
We find that the wave packet displays Bloch-like oscillations
associated with the appearance of a phase of delocalized states in
the strong correlation regime. The amplitude of oscillations
directly reflects the bandwidth of the phase and allows to measure
it. The oscillations reveal two main frequencies whose values are
determined by the structure of the underlying potential in the
vicinity of the wave packet maximum.
\end{abstract}


\section{Introduction}

Anderson localization theory describes many relevant aspects
concerning the nature of one-electron states and collective
excitations in random media~\cite{anderson,abrahams,kramer}. In
one-dimensional (1D) and two-dimensional (2D) electronic systems
with time-reversal symmetry, this theory predicts the absence of a
disorder-driven metal-insulator transition for any degree of
uncorrelated disorder. Recently, however, it has been reported that
the presence of
short~\cite{flores1,dunlap,sen,adame1,datta,evangelou1,flores2} or
long-range correlations~\cite{chico,izrailev,xiong1,Krokhin02} in
disorder acts towards the appearance of truly delocalized states in
1D Anderson models. This theoretical predictions was put forward to
account for transport properties of semiconductor superlattices with
intentional short-range correlated disorder~\cite{Bellani} and
microwave transmission spectra of single-mode waveguides with
inserted long-range correlated scatters~\cite{apl2}.

In 2D, the existence of extended states due to correlations in
disorder has been also proved. Thus, in Ref.~\cite{liu} the authors
considered a two-dimensional striped medium in the $(x,y)$ plane
with on-site correlated disorder. The on-site energies were
generated by a superposition of an uncorrelated term in the
$x$-direction and a long-range correlated contribution along
$y$-direction. It was predicted that this model displays a
disorder-driven Kosterlitz-Thouless metal-insulator transition
provided strong correlations in disorder. More recently, the effect
of an isotropic scale-free long-range correlated disorder on the
one-electron eigenstates of the 2D Anderson model has been
studied~\cite{chico2D}. To introduce long-range correlations in
\emph{both\/} $x$ and $y$ directions, the site energies were chosen
to have a power-law spectral density $S(k)\propto
1/k^{\alpha_{2D}}$, where $k$ is the magnitude of the typical
wave-vector characterizing the energy landscape roughness. The
metal-insulator transition induced by strong correlations
($\alpha_{2D}>2$) was monitored by measuring the participation
number exponent from the long-time behavior of the wave function
spatial distribution~\cite{chico2D}.

It is well known that a uniform electric field applied to a periodic
lattice causes the dynamic localization of electron wave packets and
gives rise to their oscillatory behavior, the so-called Bloch
oscillations~\cite{Bloch28}. Due to the advances in semiconductor
technology, it has become possible to monitor the Bloch oscillations
in uniform superlattices~\cite{Lyssenko97}. Remarkably, this
phenomenon is not restricted to electronic systems. Recently, the
authors of Ref.~\cite{otica2006} reported the first experimental
observation of photonic Bloch oscillations in two-dimensional
periodic systems. In spite of the fact that the periodicity of the
potential has been admitted to be the key ingredient for Bloch
oscillations to exist, it has been demonstrated recently that a 1D
Anderson model with diagonal long-range correlated disorder displays
Bloch oscillations in the strong correlation limit~\cite{prl03}. It
turned out that the period of oscillations agrees well with the one
in an ideal Bloch band, while the amplitude of oscillations is
proportional to the width of the delocalized phase, which has been
predicted to appear in the strong correlation limit~\cite{chico}.
The spectral counterpart of Bloch oscillations ---Wannier-Stark
quantization of the energy spectrum--- has been found to be also a
remarkable feature of the model~\cite{Diaz06}.

In this paper, we focus on the interplay between the
delocalization effect, arising from the long-range correlated disorder,
and the dynamic
localization, caused by an electric field acting on the system. By
solving numerically the 2D time-dependent Schr\"{o}dinger equation
for the complete Hamiltonian, we compute the behavior of an
initial Gaussian wave packet in the presence of a uniform electric
field. We found clear signatures of Bloch-like
oscillations~\cite{prl03} of the wave packet between the two
mobility edges of the phase of delocalized states. The amplitude
of the oscillatory motion of the centroid allows us to determine
the bandwidth of the delocalized phase.

\section{Model and Formalism}

We consider a 2D electron moving in a random long-range correlated
potential on a regular $N \times N$ lattice of unitary spacing and
subjected to a uniform electric field ${\bm F}$. The corresponding
tight-binding Hamiltonian reads~\cite{kramer}
\begin{eqnarray}
{\cal H} & = & \sum_{\bm m} \left(\epsilon_{\bm m}+\bm{U}\cdot
\bm{m}\right)
|{\bm m}\rangle\langle {\bm m}|
+ J \sum_{\langle \bm{mn}\rangle} \left(|{\bm
m}\rangle\langle{\bm n}|+ |{\bm n}\rangle\langle{\bm m}|\right) \ ,
\label{hamiltonian}
\end{eqnarray}
where  $|{\bm m}\rangle$ is a Wannier state localized at site
$\bm{m}=m_x{\bm e}_x+m_y{\bm e}_y$,
$\epsilon_{\bm m}$ is its energy, and
$\bm{U} = e\bm{F}$ is the energetic bias, $e$ being the electron
charge. Here ${\bm e}_x$ and
${\bm e}_y$ are the corresponding Cartesian unit vectors.
We will assume that the electric field is applied along
the diagonal of the square lattice. Then
$\bm{U} = U({\bm e}_x+{\bm e}_y)/\sqrt{2}$.
Transfer integrals are restricted to nearest-neighbor sites and
are given by $J$. Hereafter, we fix the energy scale by setting
$J=1$. The long-range correlated sequence of site energies
$\epsilon_{\bm m}$ are generated by making use of a Fourier
transform method as follows
\begin{eqnarray}
\epsilon_{\bm m}=
\sum_{k_x=1}^{N/2}\sum_{k_y=1}^{N/2}\frac{\zeta(\alpha_{2D})}{k^{\alpha_{2D}/2}}
\,\cos{\left(\frac{2\pi m_x k_x}{N}+\phi^{(x)}_{k_xk_y}\right)}
\,\cos{\left(\frac{2\pi m_y k_y}{N}+\phi^{(y)}_{k_xk_y}\right)},
\label{site_energy}
\end{eqnarray}
where $k^2=k_x^2+k_y^2$, and $\phi^{(x)}_{k_xk_y}$ and
$\phi^{(y)}_{k_xk_y}$ are $N^2/2$ independent random phases
uniformly distributed in the interval $[0,2\pi]$ and
$\zeta(\alpha_{2D})$ is a normalization constant, such that
$\langle \varepsilon_{\bm m}^2\rangle=1$. We also shift the
on-site energies in order to have $\langle \varepsilon_{\bm
m}\rangle = 0$. The Wannier amplitudes evolve in time according to
the time-dependent Schr\"{o}dinger equation which can be written
as~\cite{kramer}
\begin{eqnarray}
i\,\dot{\psi}_{\bm m}&=& \left(\epsilon_{\bm m}+ \bm{U}\cdot
\bm{m}\right)\psi_{\bm m}+\left( \psi_{{\bm m}+{\bm e}_x}+ \psi_{{\bm
m}-{\bm e}_x}+ \psi_{{\bm m}+{\bm e}_y}+
\psi_{{\bm m}-{\bm e}_y} \right)\ ,
    \label{Schrodinger}
\end{eqnarray}
with $\hbar = 1$. Having introduced the model of disorder, we
solve numerically Eq.~(\ref{Schrodinger}) to study the time
evolution of an initially Gaussian wave packet of width $\sigma$
centered at site ${\bm m}_0$
 \begin{equation}
\psi_{\bm m}(t=0)=A(\sigma)\,
\exp\left[-\,\frac{(\bm{m}-\bm{m}_0)^2}{4\sigma^2}\right]\ .
\label{Gaussian}
\end{equation}
Once Eq.~(\ref{Schrodinger}) is solved for the initial
condition~(\ref{Gaussian}), we compute the projection of the mean
position of the wave packet (centroid) along the field direction
\begin{equation}
R(t)= \frac{1}{\sqrt{2}}\, \big({\bm e}_x+
{\bm e}_y\big) \cdot \sum_{\bm m} \bm{m} \, |\psi_{\bm m}(t)|^2 \ .
\label{tools1}
\end{equation}

\section{Results and discussion}

In the absence of disorder ($\epsilon_{\bm m}=0$), all the states
are dynamically localized by the uniform electric field and the
centroid oscillates in time, revealing Bloch
oscillations~\cite{Bloch28}. The amplitude and the period of these
oscillations are estimated from semiclassical arguments (see, e.g.,
Ref.~\cite{Ashcroft76}) as $L_U = W/U$ and $\tau_B=2\pi/U$,
respectively, where $W$ is the width of the Bloch band in units of
the transfer integral $J$ ($W = 8$ in our case). The frequency of
the Bloch oscillations is given by $ \omega=2\pi/\tau_B$ and,
therefore, is equal to the bias magnitude $U$ in the chosen units.

To study joint effects of the bias and correlated disorder, we
performed numerical simulations of Eq.~(\ref{Schrodinger}). In all
simulations square lattices of size $N \times N = 250 \times 250$
were used. The initial wave packet was considered to be located in the
center of the lattice, $\bm{m}_0 = (N/2)({\bm e}_x+{\bm e}_y)$ and
its standard deviation was set to $\sigma = 1$.

\begin{figure}[ht]
\begin{center}
\includegraphics*[width=7.0cm]{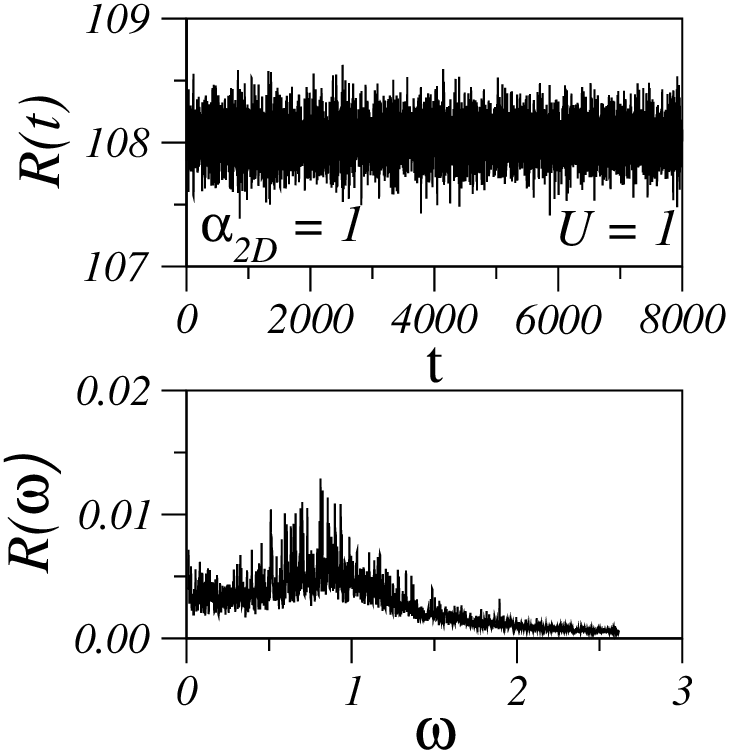}
\caption{Upper panel--Centroid dynamics of a biased wave packet ($U
= 1$) in a correlated random potential~(\ref{site_energy}) with
$\alpha_{2D}=1$. Lower panel--Fourier transform of the centroid,
averaged over 50 realization of disorder.} \label{fig1}
\end{center}
\end{figure}

In Fig.~\ref{fig1} (upper panel) we plotted the centroid time
behavior of a biased wave packet ($U=1$) in a weakly correlated
random potential ($\alpha_{2D}=1$). Recall that in the absence of
bias all the states are localized for such a value of the
correlation exponent $\alpha_{2D}$~\cite{chico2D}. As is seen,
switching on the bias does not lead to coherent oscillations of the
wave packet. A more detailed inspection of these numerical data
shows that Bloch oscillations in the weak correlation regime
($\alpha_{2D} < 2$) can be observed only for a short initial
transient. They are strongly damped by disorder and rapidly
transformed into an incoherent motion of the centroid presented in
Fig.~\ref{fig1} (upper panel). To provide a further confirmation of
this claim, we calculated the Fourier spectrum $R(\omega)$ of the
centroid $R(t)$. Figure~\ref{fig1} (lower panel) shows the result
after averaging over $50$ realizations of disorder. We observe that
the Fourier spectrum $R(\omega)$ is rather broad, suggesting that
$R(t)$ is similar to a short-range correlated noise signal with no
typical frequency.

\begin{figure}[ht]
\begin{center}
\includegraphics*[width=8.5cm]{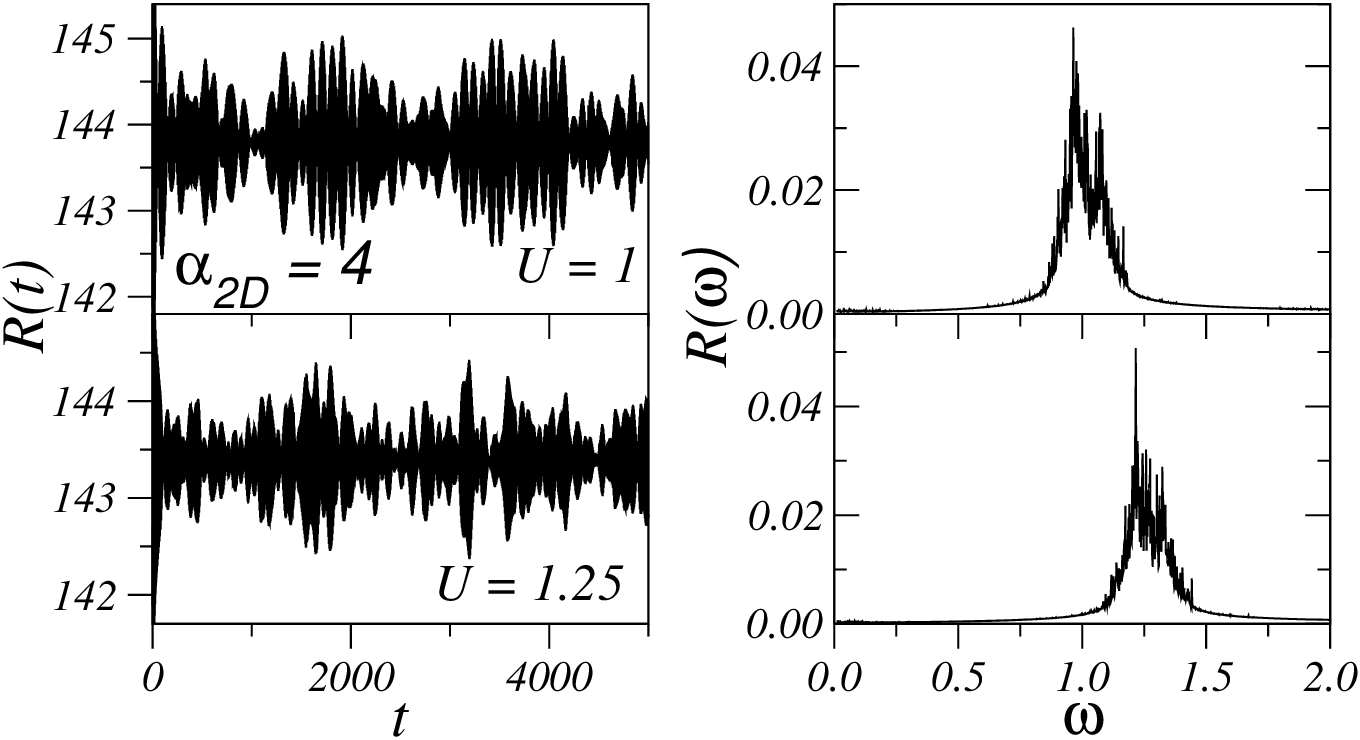}
\caption{Centroid dynamics of a biased wave packet in a correlated
random potential~(\ref{site_energy}) with $\alpha_{2D}=4$ calculated
for two bias magnitudes $U = 1$ and $U = 1.25$ (left panel). One can
see a clear signature of sustainable Bloch-like oscillations. The
corresponding Fourier spectra of the centroid, obtained by averaging
over 50 realizations of disorder, are shown in the right panel.}
\label{fig2}
\end{center}
\end{figure}

\begin{figure}[ht]
\begin{center}
\includegraphics*[width=8.5cm]{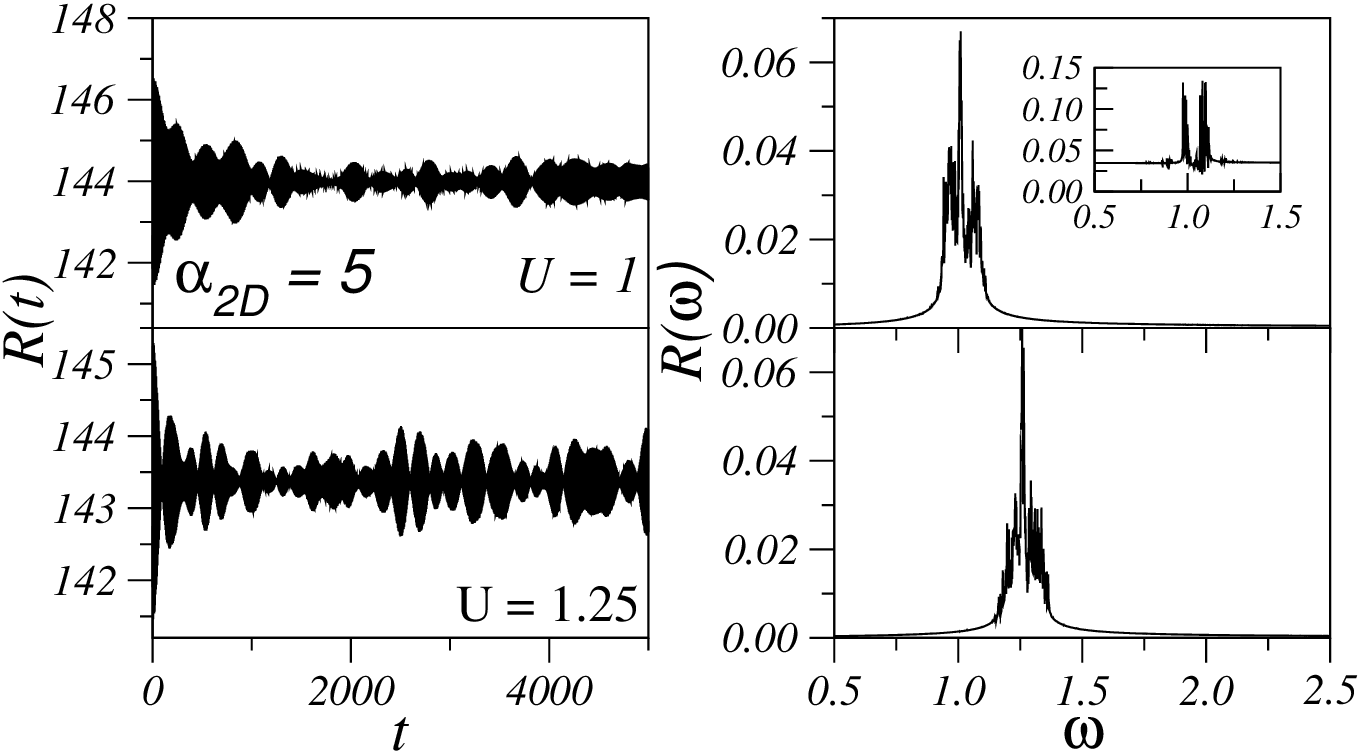}
\caption{Same as in Fig.~\ref{fig2} but for $\alpha_{2D}=5$. The
inset shows the Fourier spectrum of the centroid for a single
realization, clearly illustrating the fact that the oscillations
reveal two dominant frequencies around $\omega = U$.}
\label{fig3}
\end{center}
\end{figure}

A rather different time-domain dynamics is found in the limit of
strongly correlated disorder, $\alpha_{2D} >2$, when a phase of
extended states emerges in the center of the band in the unbiased
system~\cite{chico2D}. We will show below that these states
drastically affect the dynamics of the wave packet, that now
resembles the oscillatory motion of the electron in a biased
disorder-free lattice. In Fig.~\ref{fig2} and ~\ref{fig3} we plotted
the centroid time behavior of a biased wave packet for two values of
the correlation exponent $\alpha_{2D} = 4$ (Fig.~\ref{fig2}) and
$\alpha_{2D} = 5$ (Fig.~\ref{fig3}). Two magnitudes of the bias, $U
= 1$ and $U = 1.25$, were considered for each value of
$\alpha_{2D}$. Figures~\ref{fig2} and~\ref{fig3} clearly demonstrate
the occurrence of sustainable Bloch-like oscillations. Their
amplitudes $L_c$ are found to be $L_c \approx W_c/U$, where $W_c$ is
independent of the applied bias $U$. From the data in
Fig.~\ref{fig2} we obtain $W_c \approx 2$. This value agrees
remarkably well with the width of the band of extended states
reported in Ref.~\cite{chico2D}.

The Fourier spectra $R(\omega)$ of the centroid, computed after
averaging over $50$ realization of disorder, show a single broad
peak at about $\omega = U$. The peak frequency nicely agrees with
the one in an ideal Bloch band. Looking, however, at the Fourier
spectrum of a single realization (as shown in the inset in
Fig.~\ref{fig3}), one notices that actually $R(\omega)$ has
two narrow peaks. The frequency of these peaks fluctuates from one
realization to the other, which after averaging results in a broad
single-peaked spectral density. This splitting is not found in
1D long-range correlated potentials.

With the aim to elucidate the anomalies found, we present a
simplified model that sheds light on the origin of this doublet
structure in the Fourier spectrum of the centroid oscillations.  To
this end, we recall that the random site potential $\epsilon_{\bm
m}$ is given by the sum of harmonic terms~(\ref{site_energy}). The
amplitude of each term decreases upon increasing the harmonic
number. For sufficiently high values of $\alpha_{2D}$, the first
term in the series will be dominant, while the others are
considerably smaller. Consequently, the site potential for a given
realization represents a harmonic function, perturbed by a colored
noise (see Ref.~\cite{Diaz05}). Based on this observation, we keep
only the first term in~(\ref{site_energy}) and neglect all others in
our further argumentations. Thus, the ``random'' site potential now
is $\epsilon_{\bm m}= \zeta \cos{(2\pi m_x/N+\phi^{(x)})} \cos{(2\pi
m_y/N+\phi^{(y)}})$, where phases $\phi^{(x)}$ and $\phi^{(y)}$ can
be arbitrarily chosen.

\begin{figure}[th]
\begin{center}
\includegraphics*[width=8.5cm]{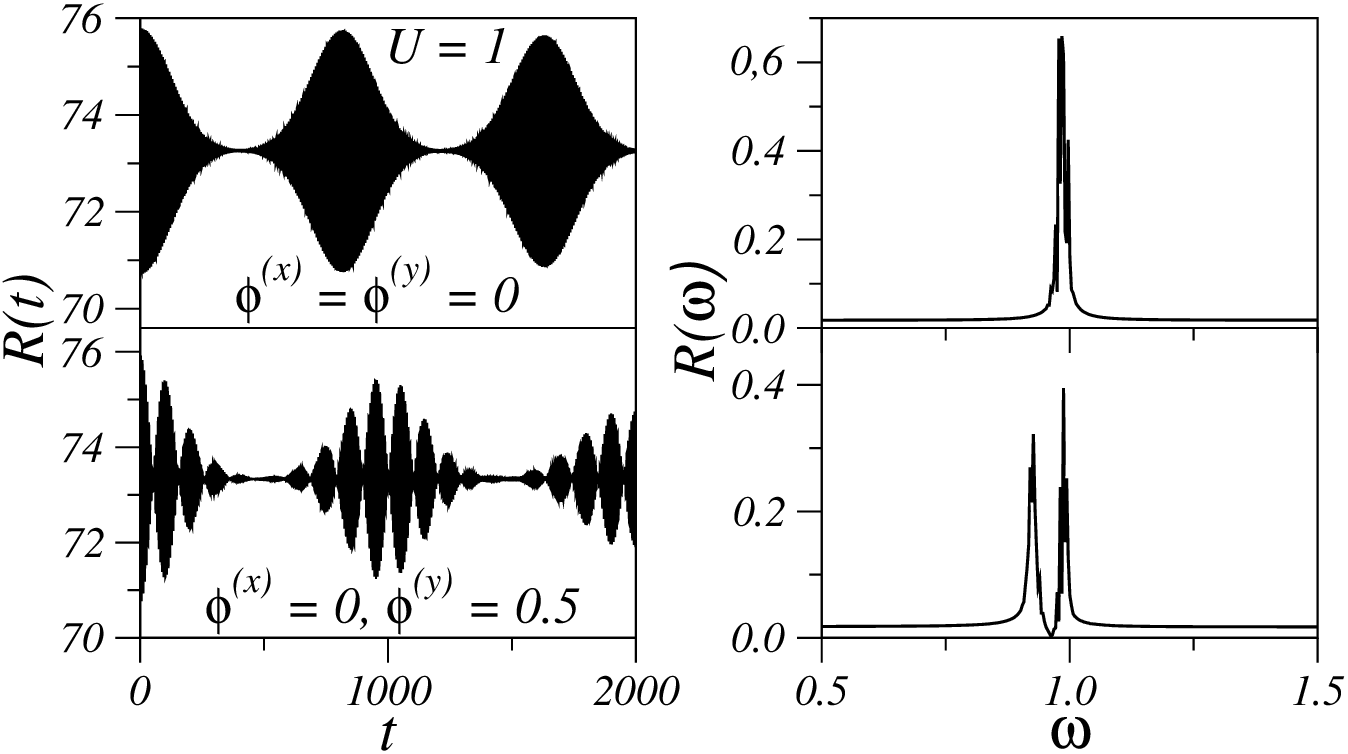}
\caption{Left panels show the centroid time behavior of a biased
    wavepacket moving in a potential $\epsilon_m =
    \zeta \cos(2\pi m_x/N + \phi_x) \cos(2\pi m_y/N + \phi_y)$ for the
    two sets of phases $\phi_x$ and $\phi_y$ indicated in the plots.
    Right panels show the corresponding centroid Fourier spectra.}
\label{fig4}
\end{center}
\end{figure}

The effective local bias is a superposition of the external one
and the gradient of the local potential, ${\bm U}_{\bm m}^{\mathrm{eff}}
={\bm U}-\nabla_{\bm m}\epsilon_{\bm m}$. At the
initial position, $\bm{m}_0 = (N/2)({\bm e}_x+{\bm e}_y)$, its components
will be given by
\begin{eqnarray}
U_{x}^{\mathrm{eff}}&=&U+(2\pi \zeta/N)\sin{\phi^{(x)}}\cos{\phi^{(y)}}\ ,\\
U_{y}^{\mathrm{eff}}&=&U+(2\pi \zeta/N)\sin{\phi^{(y)}}\cos{\phi^{(x)}}\ .
\end{eqnarray}
For $\phi^{(x)}=\phi^{(y)}$, the components of the effective bias
are identical. As the frequency of the Bloch oscillations is
proportional to the magnitude of the effective bias, only a single
dominant frequency shall be present in this case. One just observes
this in Fig.~\ref{fig4} (left-upper panel). Notice that for the
particular case plotted ($\phi^{(x)}=\phi^{(y)}=0$), the local bias
at the central position does not acquire any contribution coming
from the harmonic potential, which results in a null shift of the
oscillation frequency. However, for the general case of
$\phi^{(x)}\neq\phi^{(y)}$, the local bias components will differ by
an amount on the order of $1/N$. Therefore, Bloch oscillations will
have two dominant frequencies. This feature is exemplified in
Fig.~\ref{fig4} (left-lower panel) where we used $\phi^{(x)}=0$ and
$\phi^{(y)}=0.5$. For this case $U_x^{\mathrm{eff}}$ at ${\bm m}_0$
is not influenced by the local potential and the corresponding
oscillation frequency is not shifted. On the other hand, the typical
frequency associated with oscillations along the $y$-direction is
shifted from the bare frequency $\omega=U=1$. The two peak structure
depicted in Fig.~\ref{fig4} (left-lower panel) and in the inset of
Fig.~\ref{fig3} (right-upper panel) has, therefore, their origin in
the distinct contributions given by the gradient of the local
potential to the effective local bias. Averaging over random phases
$\phi^{(x)}$ and $\phi^{(y)}$ results in a broader Fourier spectrum
$R(\omega)$ around $\omega=U=1$ with a mean width on the order of
$1/N$.

\section{Summary and concluding remarks}

We studied the electron motion on a 2D square lattice with on-site
long-range correlated disorder in the presence of an external
uniform electric field. Long-range correlations were introduced by
using a 2D discrete Fourier method which generates an appropriate
disorder distribution with spectral density $S(k)\propto
1/k^{\alpha_{2D}}$. Solving numerically the Schrodinger equation,
the time evolution of an initial Gaussian wave packet was
investigated, aiming to find coherent Bloch oscillations. Our
results suggest that the oscillations are strongly damped in the
weak correlation limit $\alpha_{2D}< 2$, when all the states are
localized because of disorder. The motion of the wave packet on a
large time scale is chaotic in this case. The sustained oscillations
arise for $\alpha_{2D}> 2$, when a phase of the extended states
emerges at the center of the band. The amplitude of oscillations was
found to be proportional to the energy difference between the two
mobility edges of the delocalized phase~\cite{chico2D}, in good
agreement with semiclassical arguments. Thus, we arrive at one of
the principal conclusions of this work: (i)~there exist clear
signatures of the Bloch oscillations of a biased Gaussian wave
packet in the strong correlation regime ($\alpha_{2D}> 2$),
originated from the presence of the two mobility edges, and (ii)~the
oscillations exhibit two dominant frequencies because the local bias
has a contribution from the site energy potential. This is
understood using semiclassical arguments.

The richness of the predicted dynamical behavior can lead to new
electro-optical devices, based on the coherent motion of confined
electrons. We hope that the present work will stimulate
experimental activities along this direction.

\subsection{Acknowledgments}
This work was partially supported by CNPq, CAPES, and FINEP
(Federal Brazilian Agencies), and FAPEAL  (Alagoas State Agencies).
Work at Madrid was supported by MEC (Project FIS2006-01485).
V. A. M. acknowledges support from the
Stichting voor Fundamenteel Onderzoek der Materie (FOM) and from
ISTC (grant \# 2679).

\end{document}